\documentclass[preprintnumbers,amsmath,amssymb,superscriptaddress]{revtex4}

\usepackage{graphicx} 
\usepackage{bm}
\usepackage{color}
\usepackage{epstopdf}

\begin{document}

\DeclareGraphicsExtensions{.eps, .jpg}
\bibliographystyle{prsty}

\title {Infrared evidence of a Slater metal-insulator transition in NaOsO$_3$}

\author{I. Lo Vecchio}
\affiliation{Dipartimento di Fisica, Universit\`{a} di Roma "La Sapienza", Piazzale A. Moro 2, I-00185 Roma, Italy}
\author{A. Perucchi}
\affiliation{Sincrotrone Trieste, Area Science Park, I-34012 Basovizza, Trieste, Italy}
\author{P. Di Pietro}
\affiliation{Sincrotrone Trieste, Area Science Park, I-34012 Basovizza, Trieste, Italy}
\author{O. Limaj}
\affiliation{Dipartimento di Fisica, Universit\`{a} di Roma "La Sapienza", Piazzale A. Moro 2, I-00185 Roma, Italy}
\author{U. Schade}
\affiliation{Helmholtz-Zentrum Berlin f\"ur Materialien und Energie GmbH Elektronenspeicherring BESSY II, Albert-Einstein-Strasse 15, D-12489 Berlin, Germany} 
\author{Y. Sun}
\affiliation{International Center for Materials Nanoarchitectonics (MANA), National Institute for Materials Science, 1-1 Namiki, Tsukuba, Ibaraki 305-0044, Japan}
\author{M. Arai}
\affiliation{Computational Materials Science Unit, National Institute for Materials Science, 1-1 Namiki, Tsukuba, Ibaraki 305-0044, Japan}
\author{K. Yamaura}
\affiliation{Superconducting Properties Unit, National Institute for Materials Science, 1-1 Namiki, Tsukuba, Ibaraki 305-0044, Japan}
\author{S. Lupi}
\affiliation{CNR-IOM and Dipartimento di Fisica, Universit\`a di Roma "La Sapienza", Piazzale A. Moro 2, I-00185, Roma, Italy}
\date{\today}

\begin{abstract}
The magnetically driven metal-insulator transition (MIT) was predicted by Slater in the fifties. Here a long-range antiferromagnetic (AF) order can open up a gap at the Brillouin electronic band boundary regardless of the Coulomb repulsion magnitude. However, while many low-dimensional organic conductors display evidence for an AF driven MIT, in three-dimensional (3D) systems the Slater MIT still remains elusive. We employ terahertz and infrared spectroscopy to investigate the MIT in the NaOsO$_3$ 3D antiferromagnet. From the optical conductivity analysis we find evidence for a continuous opening of the energy gap, whose temperature dependence can be well described in terms of a second order phase transition. The comparison between the experimental Drude spectral weight and the one calculated through Local Density Approximation (LDA) shows that electronic correlations play a limited role in the MIT. All the experimental evidence demonstrates that NaOsO$_3$ is the first known 3D Slater insulator.

\end{abstract}

\maketitle

In the last sixty years considerable effort has been devoted to the study of the MIT in transition metal oxides \cite{Imada-98,Perucchi-rev}. In their fundamental papers Mott \cite{Mott-49} and Hubbard \cite{Hubbard-63} proposed that the electron-electron interaction (\textit{U}) is the key ingredient of the MIT. Electronic correlations may strongly renormalize the kinetic energy (\textit{t}) of charge-carriers leading to the formation of a gap at the Fermi energy and then to an insulating state regardless of magnetic correlations. Temperature and/or pressure may increase the $t/U$ ratio, thus inducing a transition (Mott-Hubbard transition) from the insulating to the metallic state. V$_2$O$_3$\cite{Lupi-10}, Ni(S,Se)$_2$\cite{Perucchi-09} and NdNiO$_3$\cite{Basov-11} are considered textbook examples of Mott-Hubbard materials. On the other hand, Slater \cite{Slater-51} suggested a different mechanism for the MIT: he stated that antiferromagnetic order alone could open up a gap. The establishment of a commensurate AF order induces an opposite potential on each nearest electron site which doubles the magnetic unit cell and cuts the first Brillouin zone. This results in a splitting of the occupied bands and, in case of half-filling, in a gap for charge excitations.
So far the only evidence of magnetically driven MITs was found on low dimensional organic compounds as Bechgaard salts \cite{Gruner-94, DeGiorgi-96, Vescoli-99}, while the 3D candidates Cd$_2$Os$_2$O$_7$ \cite{Mandrus-01,Padilla-02,Yamaura-12} and Ln$_2$Ir$_2$O$_7$ \cite{Matsuhira-11} could not be classified as Slater insulators due to magnetic frustration in their pyrochlore lattice \cite{Reading-01}. Experimental indications of a Slater transition have never come up until last year \cite{Calder-12}. By using neutron and x-ray scattering, Calder \textit{et al.} have shown that in the perovskite NaOsO$_3$ the MIT at T$_{MIT}$=410 K, previously discovered on the basis of transport and magnetic measurements \cite{Shi-09}, is concomitant with the onset of long-range commensurate three-dimensional antiferromagnetic order. As a perovskite with octahedral environment Os$^{5+}$O$_6$,  NaOsO$_3$ shows a half-filled 5\textit{d}$^3$ electronic configuration. In addition, 5\textit{d} orbitals are far more spatially extended than those in 3\textit{d} systems, thus electron-electron interactions is expected to  play a minor role in this compound, as also recently argued from theoretical calculations \cite{Du-12, Pickett-13}.
In this work we probe the low-energy electrodynamics of NaOsO$_3$ across the MIT by terahertz and infrared spectroscopy. From an analysis of the optical spectral weight we establish that NaOsO$_3$ is a weakly correlated material \cite{Basov-09}. We also show that, at variance with a Mott-Hubbard MIT \cite{Baldassarre-08, Lupi-10, Perucchi-09, Basov-11}, the optical conductivity does not vanish at T$_{MIT}$ as the charge gap opens up in a continuous way in agreement with the second order character of the MIT. Those experimental results definitively clarify the nature of the three-dimensional MIT in NaOsO$_3$ in terms of the Slater mechanism.

\section*{Results}
The reflectance R($\omega$) of NaOsO$_3$ is shown in  Fig. 1 in the 0-1000 cm$^{-1}$ frequency range. At 450 K it shows a metallic response, approaching unity at zero frequency. After crossing T$_{MIT}$ this metallic behavior depletes progressively and, in the antiferromagnetic insulating phase, two complex phononic structures start to appear as a consequence of the reduced screening. At 5 K there is a well visible double phonon peak with characteristic frequencies around 300 cm$^{-1}$ and 330 cm$^{-1}$ and another phonon resonance at 650 cm$^{-1}$. Moreover, R($\omega$) shows a rising tail for $\omega\rightarrow0$ related to a low-frequency mode centered around 20 cm$^{-1}$ (see discussion below). In the inset of Fig. 1 we show instead the reflectance over the entire measured range in the metallic (450 K) and in the insulating state (5 K). It is well evident that the MIT determines a strong modification of the electronic properties of NaOsO$_3$ over a frequency scale up to nearly 10000 cm$^{-1}$.

The main panel of Fig. 2 shows the optical conductivity $\sigma_{1}(\omega)$ on a linear scale in the 0-10000 cm$^{-1}$ frequency range as obtained from reflectance data by Kramers-Kronig relations. The same quantity is represented in the inset on a logarithmic scale for selected temperatures above and below T$_{MIT}$. The symbols on the vertical axis indicate the dc conductivity values calculated from resistivity data measured on samples coming from the same batch \cite{Shi-09}. At all temperatures there is a good agreement between the zero frequency limit of $\sigma_{1}(\omega)$ and the measured $\sigma_{dc}$. 

The optical conductivity at 450 K shows a metallic behavior with a broad pseudo plasma-edge around 12000 cm$^{-1}$ which separates the low energy excitations from a huge interband transition around 20000 cm$^{-1}$. This absorption band is mainly associated to charge-transfer excitations among Os 5d and O 2p states \cite{Shi-09, Du-12}. Below T$_{MIT}$ the metallic conductivity sharply decreases in the far-infrared through a transfer of spectral weight (SW) to a mid-infrared (MIR) band centered around 3000 cm$^{-1}$. The low-frequency SW depletion is nearly exhausted at 200 K where the MIR is located at about 4000 cm$^{-1}$. 

A broad feature centered around 22 cm$^{-1}$ can be seen in the insulating phase between 5 and 200 K and it is probably hidden at higher-T by the free carrier background. We associate this peak to an antiferromagnetic resonance. Indeed, the insulating phase of NaOsO$_3$ corresponds to a G-type AF configuration \cite{Calder-12} where the spins are oriented along the \textit{c}-axis. The antiferromagnetic mode corresponds to a precession of spins along $c$ induced by some degree of magnetic anisotropy \cite{Kittel-52, Sievers-63}.  

The loss of SW in the far-infrared mirrors the opening of an optical gap E$_g$ already distinguishable at 380 K, whose size increases for decreasing T. 
One can then extract E$_g$ from data in Fig. 2 and compare its temperature dependence with that expected for a second order (Slater) phase transition. 
The optical conductivity in the insulating AF phase can be described through the following equations:

\begin{equation}
\sigma_{1}(\omega)=m\omega+q \hspace{0.5cm}\mbox{for}\hspace{0.2 cm} \omega\leq E_g
\label{linear}
\end{equation}

and 

\begin{equation}
\sigma_{1}(\omega)=A[\omega-E_g(T)]^\alpha \hspace{0.5cm}\mbox{for}\hspace{0.2 cm} \omega\geq E_g
\label{parabolic}
\end{equation}

\noindent 
Here, the linear term describes the low-frequency background which is mainly due to thermally excited charge-carriers across the insulating gap (see below).
For higher frequencies instead, we assumed a behavior of $\sigma_{1}(\omega)$ similar to that of a semiconductor in the presence of direct band to band transitions \cite{Burns}. The resulting curves, reported as dashed lines in the inset of Fig. 3, nicely fit the rising edge of $\sigma_{1}(\omega)$ at all temperatures with $\alpha=\frac{1}{2}$. An estimation of the gap E$_g$ is then obtained through the intersection between Eq. \ref{linear} and Eq. \ref{parabolic}, and its temperature behavior, plotted as empty circles in Fig. 3, resembles that of a Bardeen-Cooper-Schrieffer (BCS) function in good agreement with the second order character of the MIT. Let us mention that similar gap values can also be achieved by substituting the linear conductivity (Eq. \ref{linear}) with an overdamped Drude term. Therefore the error bars in Fig. 3 take into account the small variation of E$_g$(T) in terms of different extracting methods and the smearing effect in the conductivity edge due to the temperature.

A BCS-like analytic expression for E$_g(T)$ can be written as \cite{Burns}:

\begin{equation}
\frac{E_g(T)}{E_g(0)}=tanh \frac{E_g(T)T_{MIT}}{E_g(0)T}
\label{BCS}
\end{equation}

\noindent This function well describes data in Fig. 3 furnishing a value of E$_g$(0)=825 $\pm$ 25 cm$^{-1}$ and a T$_{MIT}$=400 $\pm$ 10 K in fair agreement with the value (T$_{MIT}$=410 K) obtained through transport and magnetic measurements \cite{Calder-12, Shi-09}. Moreover, E$_g(0)$/k$_B$T$_{MIT}$=3.0$ \pm$ 0.1, thus suggesting a weak coupling regime in agreement with a Slater scenario. 

As mentioned before, the gap opening corresponds to a loss of SW in the infrared which is transferred to higher frequencies. The energy scale involved in this process can be estimated from Fig. 4 where the optical spectral weight is reported. It can be written as follows:

\begin{equation}
SW(\omega_c,T)=\int_{0}^{\omega_c}\sigma_1(\omega,T)\mathrm{d}\omega
\label{SW1}
\end{equation}
 
\noindent  and it is proportional to the number of carriers taking part to the optical absorption up to a cutoff frequency $\omega_c$. The SW is nearly conserved for $\omega_c\simeq$15000 cm$^{-1}$. This energy scale is well captured by the LDA+U calculation reproducing, for a G-type antiferromagnetic order, the insulating state for a moderate value of \textit{U}$\sim$1-2 eV \cite{Calder-12, Du-12, Pickett-13}.  

The SW redistribution can be analyzed in detail by fitting $\sigma_{1}(\omega)$ through a multi-component Drude-Lorentz (D-L) model. The complex optical conductivity is written in terms of a Drude contribution and five Lorentzian curves as:

\begin{equation}
\tilde{\sigma}(\omega)=\frac{\omega_P^2\tau}{4\pi(1-i\omega\tau)}+\frac{\omega}{4\pi i}\sum_j\frac{S_j^2}{\omega_j^2-\omega^2-i\omega\gamma_j}
\label{DrudeLorentz}
\end{equation}

In the Drude term $\omega_P$ is the plasma frequency, $\tau$ is the scattering time, while the Lorentz oscillators are peaked at finite frequencies $\omega_j$ with strength $S_j$ and width $\gamma_j$. We used a Lorentzian oscillator both for the mid-infrared band and for the charge-transfer transition around 20000 cm$^{-1}$. The phonon absorption is described in terms of three Lorentzian components. 
Examples of D-L fits are represented in the main panel of Fig. 2 both at 450 K (metallic phase) and 5 K (insulating phase). From the fit we have extracted $\omega_P^2$ whose values vs. T are shown in the inset of Fig. 4 as open diamonds. One can observe a drastic collapse of $\omega_P^2$ at the MIT. Below T$_{MIT}$ the residual Drude weigth is mainly determined by the charge-carriers thermally excited above the insulating gap. Indeed $\omega_P^2$ can be fitted by an exponential T-dependence $\omega_P^2(T)$=$Ae^{-E_g(T)/2k_BT}$ (red symbols) with $A=4.6\cdot10^7$ cm$^{-2}$, nearly corresponding to the square of the plasma frequency at 410 K. 
By employing the gap values in Fig. 3, this exponential equation  properly reproduces the experimental points from 200 K, where the Drude contribution is still zero, to 380 K where the gap is nearly closed.  

Even more importantly the plasma frequency at 450 K can be used for a quantitative determination of the degree of electronic correlation in NaOsO$_3$.
Indeed the Drude plasma frequency is related to the experimental kinetic energy of mobile carriers by the relation $K_{exp}\sim\omega_P^{2}/8$. Due to correlation effects this value is renormalized (reduced) with respect to the kinetic energy provided by Local Density Approximation calculations $K_{LDA}$ \cite{Basov-09}. In a Mott-Hubbard insulator $K_{exp}$=0 as charge-carriers are completely localized due to Coulomb correlation, at variance with the band-structure results predicting a metallic response with a finite $K_{LDA}$. Conversely, for conventional metals like gold and copper, the ratio $K_{exp}/K_{LDA}\sim$1, indicating the scarce relevance of electronic correlations. From the Drude-Lorentz fit at 450 K shown in the main panel of Fig. 2 we estimate $\omega_P\sim$1.6 eV; LDA calculations provide $\omega_P\mbox{(LDA)}=2.1$ eV instead. From these data we achieve a ratio $K_{exp}/K_{LDA}\sim$0.6, thus locating NaOsO$_3$ at the border between correlated metals and uncorrelated systems suggesting a relatively weak electronic correlation.

%<<<<<<<<<<<<<<<<<FIG1>>>>>>>>>>>>>>>>>>>>>>>>>>>>>>>>>>>>>>>>>>>>>>>>>>>>>----------
\begin{figure}[h]
\begin{center}
\leavevmode
\includegraphics [width=9cm]{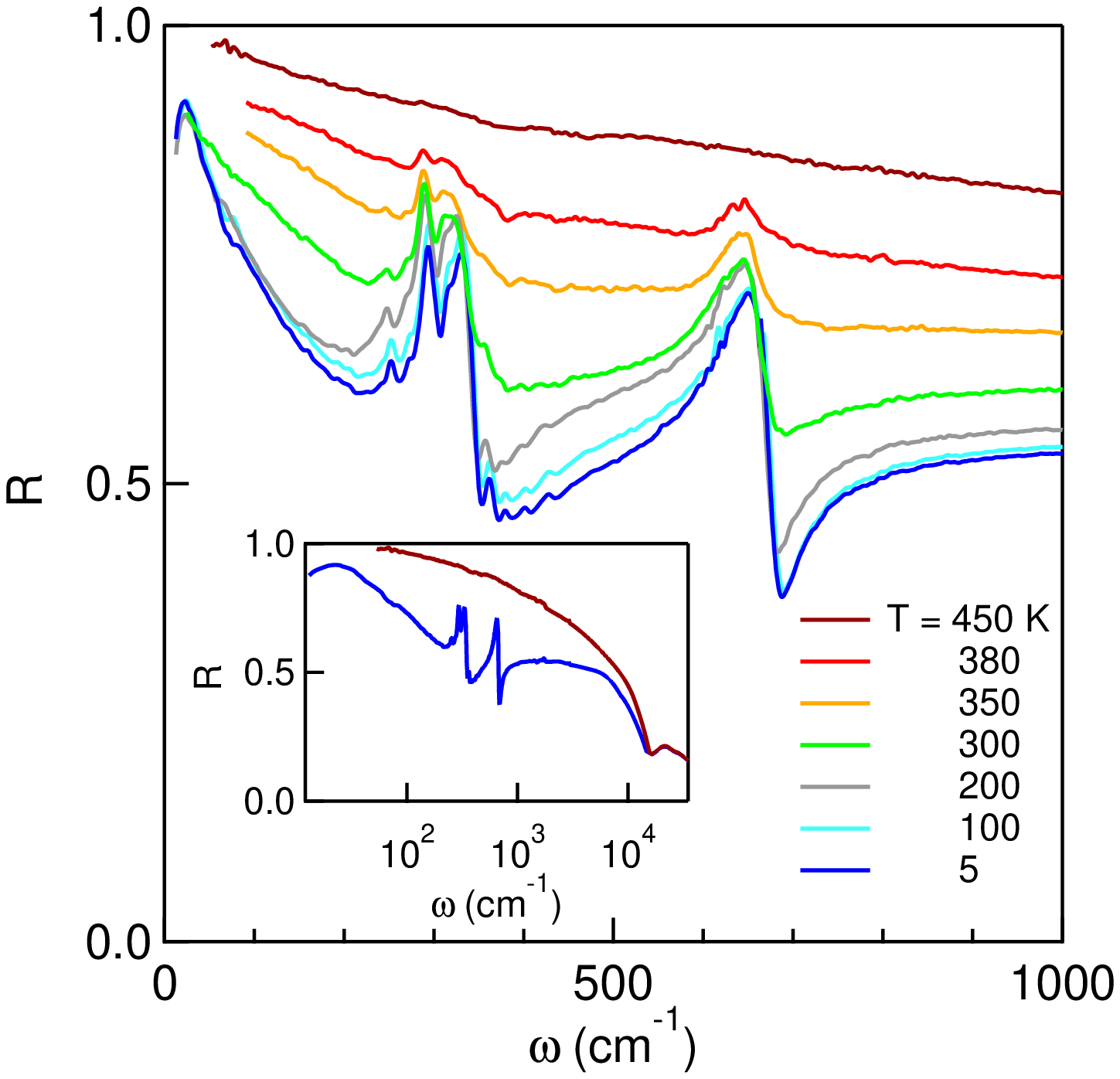}  
\end{center}  
\end{figure}
%<<<<<<<<<<<<<<<<<END FIG1>>>>>>>>>>>>>>>>>>>>>>>>>>>>>>>>>>>>>>>>>>>>>>>>>>>

%<<<<<<<<<<<<<<<<<FIG 2>>>>>>>>>>>>>>>>>>>>>>>>>>>>>>>>>>>>>>>>>>>>>>>>>>>>>>>
\begin{figure}[h]
\begin{center}
\leavevmode
\includegraphics [width=9cm]{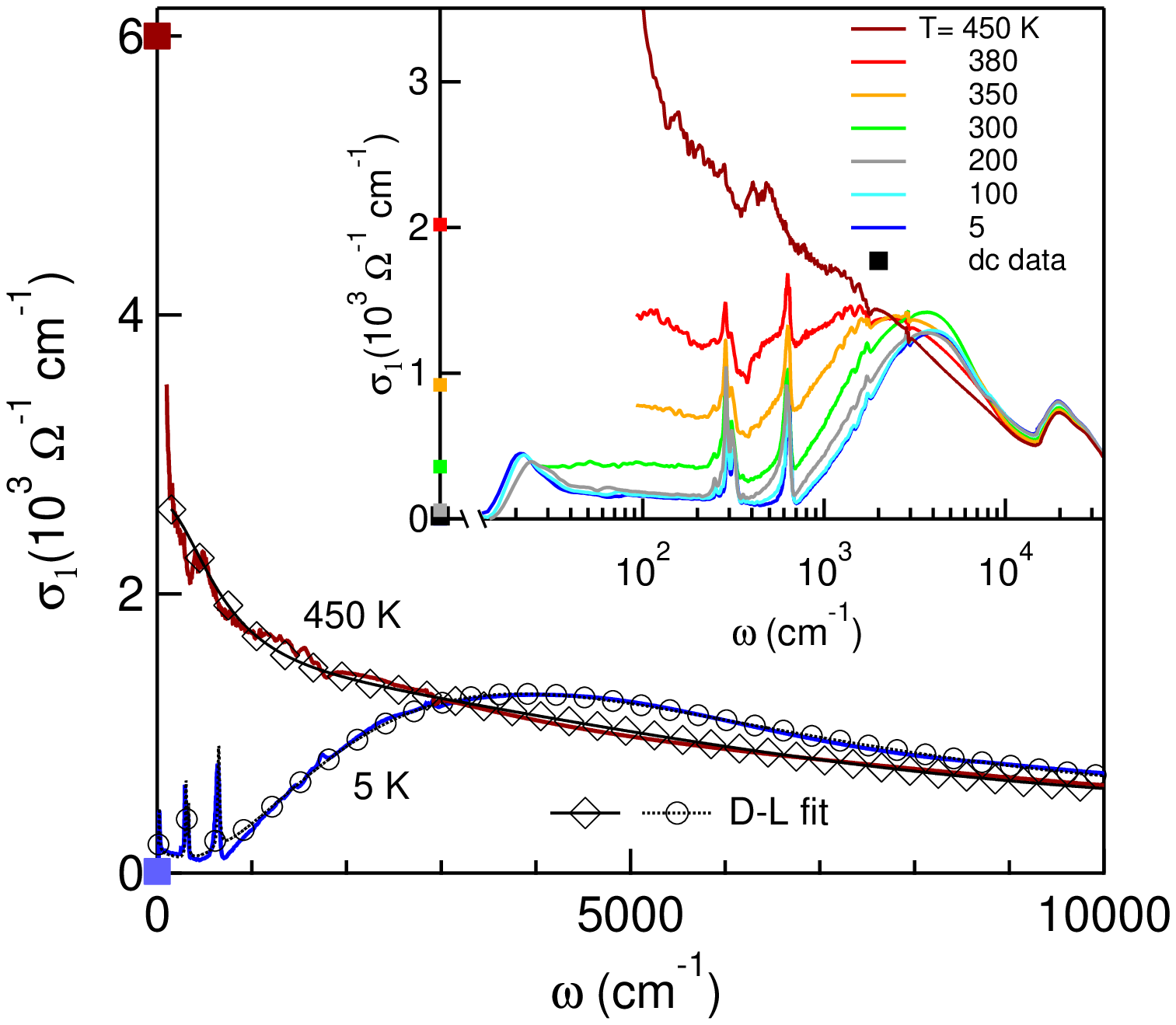}  
\end{center}
\end{figure}
%<<<<<<<<<<<<<<<<<END FIG 2>>>>>>>>>>>>>>>>>>>>>>>>>>>>>>>>>>>>>>>>>>>>>>>>

%<<<<<<<<<<<<<<<<<FIG 3>>>>>>>>>>>>>>>>>>>>>>>>>>>>>>>>>>>>>>>>>>>>>>>>>>>>>>>
\begin{figure}[h]
\begin{center}
\leavevmode
\includegraphics [width=9cm]{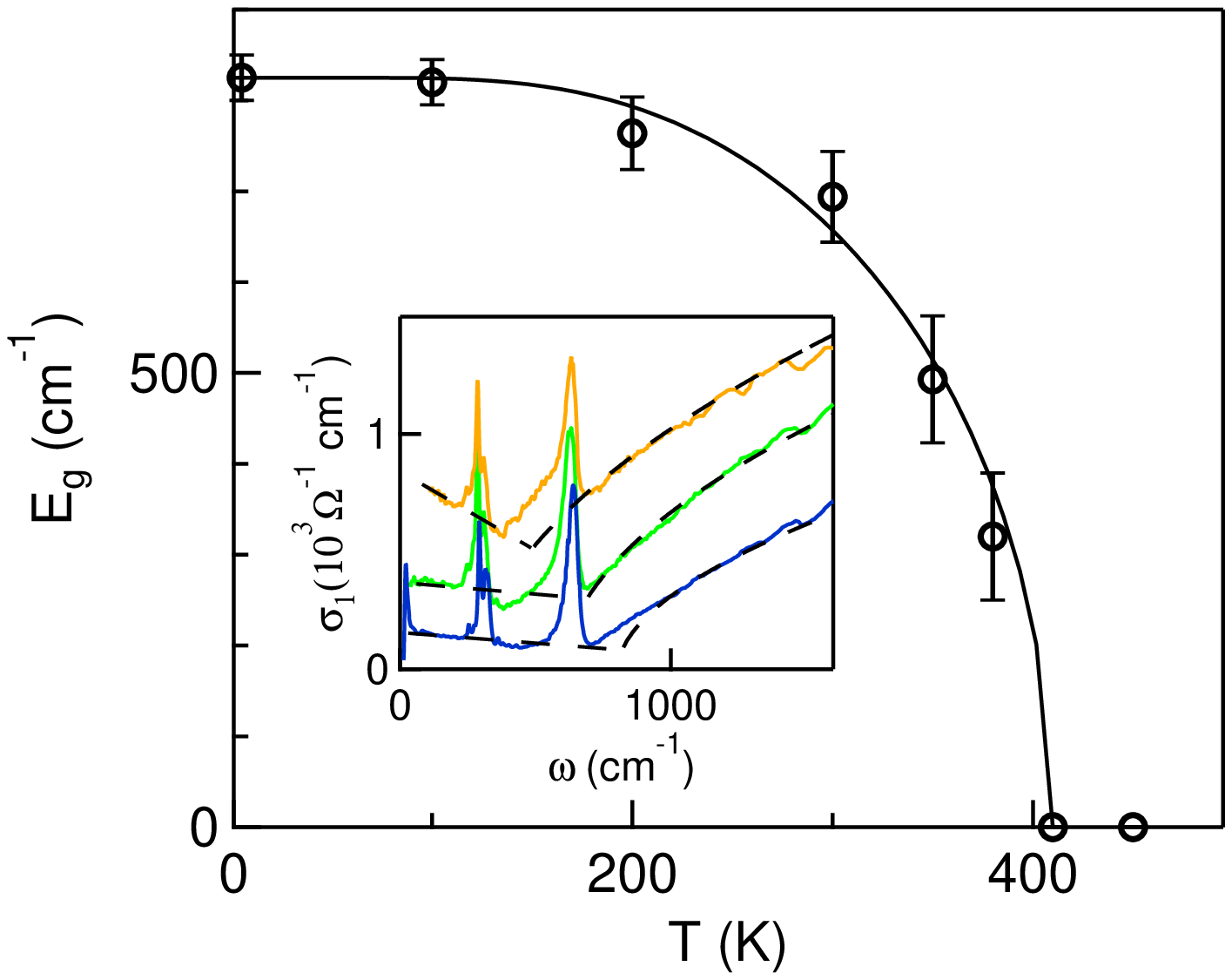}  
\end{center}
\end{figure}
%<<<<<<<<<<<<<<<<<END FIG 3>>>>>>>>>>>>>>>>>>>>>>>>>>>>>>>>>>>>>>>>>>>>>>>>>>>>>

%<<<<<<<<<<<<<<<<<FIG 4>>>>>>>>>>>>>>>>>>>>>>>>>>>>>>>>>>>>>>>>>>>>>>>>>>>>>>>
\begin{figure}[h]
\begin{center}
\leavevmode
\includegraphics [width=9cm]{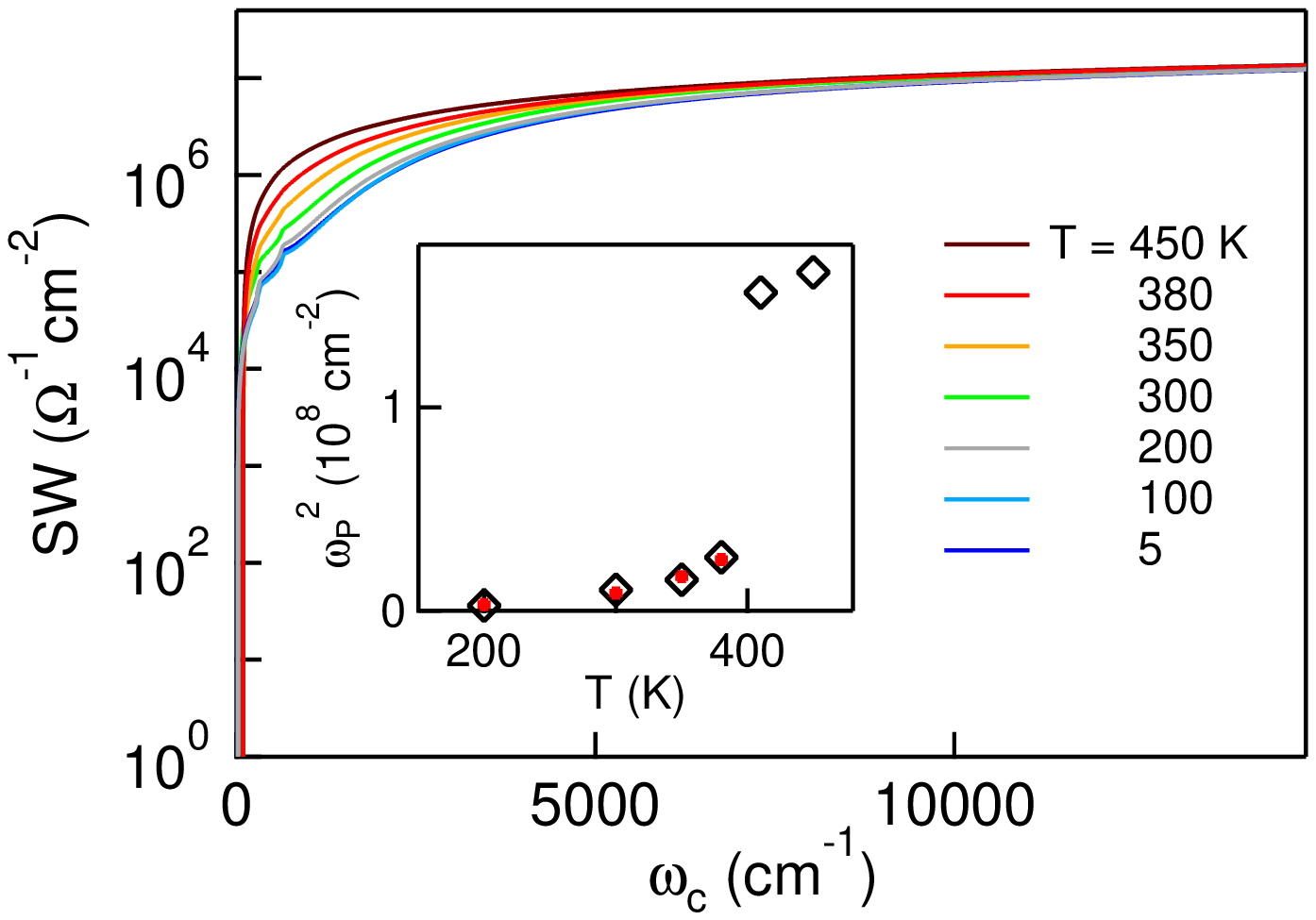}  
\end{center}
\end{figure}
%<<<<<<<<<<<<<<<<<END FIG 4>>>>>>>>>>>>>>>>>>>>>>>>>>>>>>>>>>>>>>>>>>>>>>>>>>>>>

\section*{Discussion}
NaOsO$_3$ shows a half-filled 5\textit{d}$^3$ electronic configuration. Since 5\textit{d} orbitals are spatially more extended than those in 3\textit{d} and 4\textit{d} systems, theoretical calculations argue that electron-electron interactions should play a minor role in the low-energy electrodynamics suggesting a Slater origin of the MIT \cite{Du-12, Pickett-13}. In this paper we provide robust experimental evidence for the Slater mechanism. Indeed, the $K_{exp}/K_{LDA}$ ratio, which has been demonstrated to well estimate the electronic correlation degree in many materials, gets a value $K_{exp}/K_{LDA}\sim$0.6. This firstly suggests that NaOsO$_3$ lies near the band-like itinerant regime \cite{Basov-09}. Secondly, as demonstrated in Fig. 3, the gap develops in a continuous way in NaOsO$_3$, by following a BCS-like temperature dependence. Conversely, in Mott-Hubbard materials as V$_2$O$_3$, Ni(S,Se)$_2$ and NdNiO$_3$, a gap abruptly appears while crossing T$_{MIT}$\cite{Baldassarre-08}. Moreover, both the E$_g(0)$/k$_B$T$_{MIT}$=3.0 $\pm$ 0.1 ratio and the gap temperature dependence  are in good agreement with the second order character of a Slater MIT. Finally, the concomitance of the MIT with the AF ordering, as measured through neutron scattering, suggests that magnetism is the main cause for the MIT. All these findings clarify the MIT in terms of the Slater mechanism, showing that NaOsO$_3$ is the first known 3D Slater insulator.

\section*{Methods}
{\small
{\bf Sample preparation.} A high-density polycrystalline pellet sample of NaOsO$_3$ was synthesized starting from Na$_2$O$_2$ and OsO$_2$ powders through a high pressure technique (see Ref. \onlinecite{Shi-09}). A powder x-ray diffraction study using Cu $K\alpha$ radiation confirmed the absence of impurities. 

{\bf Optical measurement.} We performed near-normal reflectance measurements on an accurately polished sample using a Michelson interferometer in the frequency range from 10 cm$^{-1}$ (this limit varies for some temperatures) to 15000 cm$^{-1}$. Terahertz and subterahertz ($\omega<$50 cm$^{-1}$) measurements were performed using radiation produced at the SISSI \cite{SISSI-07} and IRIS \cite{IRIS-03} infrared beamlines at Elettra and Bessy-II synchrotrons respectively.  A gold or silver (depending on the spectral range) coating was evaporated \textit{in situ} over the sample surface and used as a reference. The optical conductivity was calculated by Kramers-Kronig transformations. Low-frequency reflectance data were extrapolated with standard methods (Hagen-Rubens or constant lines) taking into account the resistivity dc values measured in samples of the same batch. A high-frequency tail \cite{Padilla-02} was instead merged to the data above 15000 cm$^{-1}$. 

{\bf LDA calculation.} 
The $K_{LDA}$ was obtained from first-principles calculations within Local Density Approximation (LDA) of density functional theory. The
electronic structure was calculated from full-potential augmented plane wave (APW) methods with the WIEN2k package \cite{Blaha-01}. The spin-orbit
interaction was included within self-consistent calculations.  The lattice constants and atomic positions were taken from experimental
values \cite{Shi-09}. The Muffin-tin radii were chosen as 2.0, 1.8 and 1.7 a.u. for Na, Os, and O, respectively. The cutoff wave number K in
interstitial region was set to RK = 8, where R is the smallest atomic radius, i. e., 1.7 a.u. for O. The integration over Brillouin zone was performed by a tetrahedron method with up to 1960 k-points in an irreducible zone. The optical conductivity and plasma frequency were estimated from the momentum matrix elements as described in Ref. \onlinecite{Ambrosch-06}. Since NaOsO$_3$ has orthorhombic symmetry with \textit{Pnma} space group, the optical conductivity tensors are characterized by three diagonal elements $\sigma_{xx}$, $\sigma_{yy}$, and $\sigma_{zz}$. Accordingly, we obtained plasma frequency for each direction as $\omega_{P xx}=2.2$ eV, $\omega_{P yy}=1.8$ eV, $\omega_{P zz}=2.4$ eV.  The averaged plasma frequency $\omega_P(\mbox{LDA})=(\omega_{P xx} + \omega_{P yy} + \omega_{P zz})/3$
was estimated as 2.1 eV.}

\section{Authors Contributions} 
Y.S. and K.Y., fabricated and characterized NaOsO$_3$ samples. I.L.V., A.P., P.D.P., O.L. U.S. and S.L. carried out the terahertz experiments and data analysis. M.A. was responsible of the LDA calculations. S.L., was responsible for the planning and the management of the project with inputs from all the co-authors, especially from I.L.V., A.P. and K.Y. All authors extensively discussed the results and the manuscript that was written by I.L.V., A.P. and S.L. 
 
\section{Additional Information}
The authors declare no competing financial interests. Correspondence and requests for materials should be addressed to S. L. (stefano.lupi@roma1.infn.it)

\begin{widetext}

FIG 1: {\bf The near-normal incidence reflectance of NaOsO$_3$ is reported at selected temperatures from 5 K to 450 K on a linear scale in the 0-1000 cm$^{-1}$ frequency range. In the inset data are shown in the insulating (5 K) and in the metallic (450 K) states on a log scale in the whole frequency range. The MIT temperature is 410 K.}\\

FIG 2: {\bf Optical conductivity of NaOsO$_3$ at 5 K and 450 K
on a linear frequency scale as obtained from Kramers-Kronig transformations. Symbols on the vertical axis stand for dc values of the conductivity, calculated from transport measurements.
The inset shows the same data on a log scale at selected temperatures. The Drude-Lorentz fitting results (see text) are indicated by open symbols.}\\

FIG 3: {\bf Magnitude of the optical gap with its fit to Eq.(\ref{BCS}). The extrapolations of the optical conductivity based on Eq. (1) and (2) are shown in the inset as dashed curves at T=350, 300, 5 K (from top to bottom).}\\

FIG 4: {\bf Optical spectral weight vs cutoff frequency $\omega_c$ at selected temperatures. The SW is recovered at nearly 15000 cm$^{-1}$. Inset: the squared plasma frequency vs T behavior is the result of the Drude-Lorentz fit analysis of the optical conductivity. Values calculated from the relation $Ae^{-E_g(T)/2k_BT}$ are indicated by red symbols.}\\

\end{widetext}

\end{document}